\documentclass{iau} 
\usepackage{graphicx}

\title[Evolution of bright PNe] 
{Several evolutionary channels for bright planetary nebulae}

\author[Michael Richer \& Marshall McCall]   
{Michael G. Richer$^1$
 \and Marshall L. McCall$^2$
}

\affiliation{$^1$Instituto de Astronom\'\i a, Universidad Nacional Aut\'onoma de M\'exico, Apartado Postal 106, 22800 Ensenada, Baja California, M\'exico  email: {\tt richer@astrosen.unam.mx}\\

$^2$Department of Physics and Astronomy, York University, Toronto, Ontario L3T 3R1, Canada  \\email: {\tt mccall@yorku.ca}}

\pubyear{2015}
\volume{xxx}  
\jname{Title of your IAU Symposium}
\editors{A.C. Editor, B.D. Editor \& C.E. Editor, eds.}
\begin{document}

\maketitle

\begin{abstract}
The populations of bright planetary nebulae in the discs of spirals appear to differ in their spectral properties from those in ellipticals and the bulges of spirals.  The bright planetary nebulae from the bulge of the Milky Way are entirely compatible with those observed in the  discs of spiral galaxies.  The similarity might be explained if the bulge of the Milky Way evolved secularly from the disc, in which case the bulge should be regarded as a pseudo-bulge.
\end{abstract}

\firstsection 
\section{Introduction}

Planetary nebulae (PNe) are the penultimate evolutionary stage of stars in the 1-8 M$_{\odot}$ mass range.  Bright PNe (large absolute luminosities in the [O {\sc iii}] $\lambda$5007 line) are useful for studying aspects of stellar and galactic evolution in the nearby universe.  The bright end of the luminosity function of PNe in [O {\sc iii}] $\lambda$5007 (PNLF) has been used as a secondary distance indicator for galaxies of all morphological types (e.g., \cite[Ciardullo et al. 1989]{ciardulloetal1989}).  While the physics behind the constancy of the PNLF peak luminosity is not clearly understood, empirical evidence indicates that the brightest PNe in all galaxies reach similar luminosities, have nebular shells whose kinematics are similar, and arise from stellar progenitors that have undergone similar nucleosynthetic processes (\cite[Ciardullo et al. 2002; Richer \& McCall 2008; Richer et al. 2010]{ciardulloetal2002, richermccall2008, richeretal2010}).  The internal kinematics evolve with time, but not necessarily in the same way in all galaxies.  It is tempting to suppose that the same stellar progenitors produce the brightest PNe in all galaxies, but it is unclear whether this is feasible since the dominant stellar populations vary from galaxy to galaxy, both in mass and metallicity.  

PNe are dynamical systems.  The state of both the central star and the nebular shell change substantially with time.   Models predict that, initially, the luminosity in H$\beta$ increases very rapidly, but that the [O{\sc iii}] $\lambda$5007 luminosity overtakes it some time later (e.g., \cite[Sch\"onberner et al. 2007]{schonberneretal2007}).  As a result, both the H$\beta$ luminosity and 5007/H$\beta$ ratio vary with time, probing to some extent the luminosity and temperature, respectively, of the central star.  How the 5007/H$\beta$ ratio varies as a function of the H$\beta$ luminosity will depend upon the details of the central star and nebular shell and how they vary with time (e.g., the nebular mass and its distribution, the opacity of the nebular shell to ionizing photons, and the relative evolutionary rates of the central star and nebular shell).

\begin{figure}[]
\begin{center}
 \includegraphics[width=0.473\columnwidth]{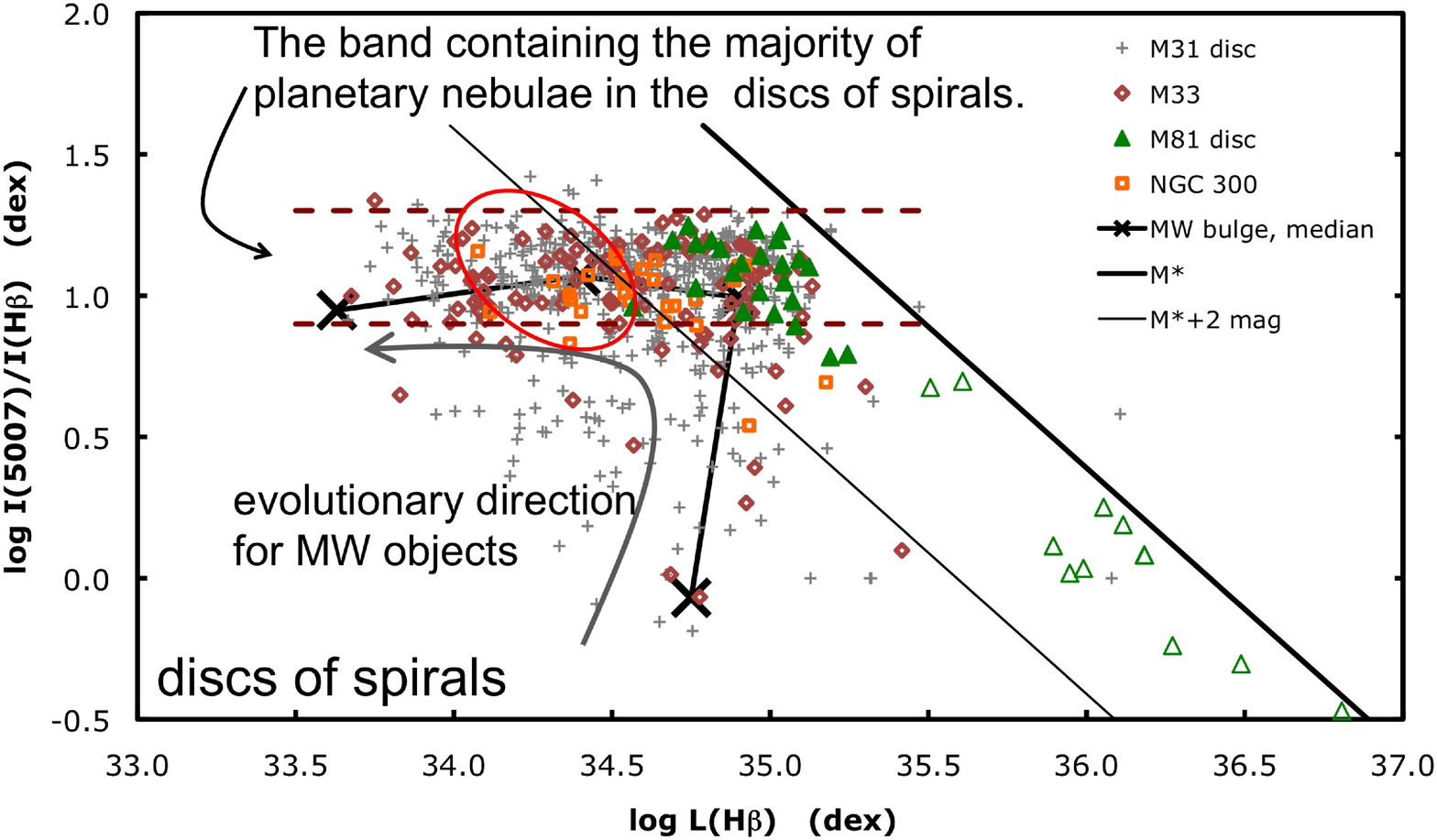} 
 \includegraphics[width=0.487\columnwidth]{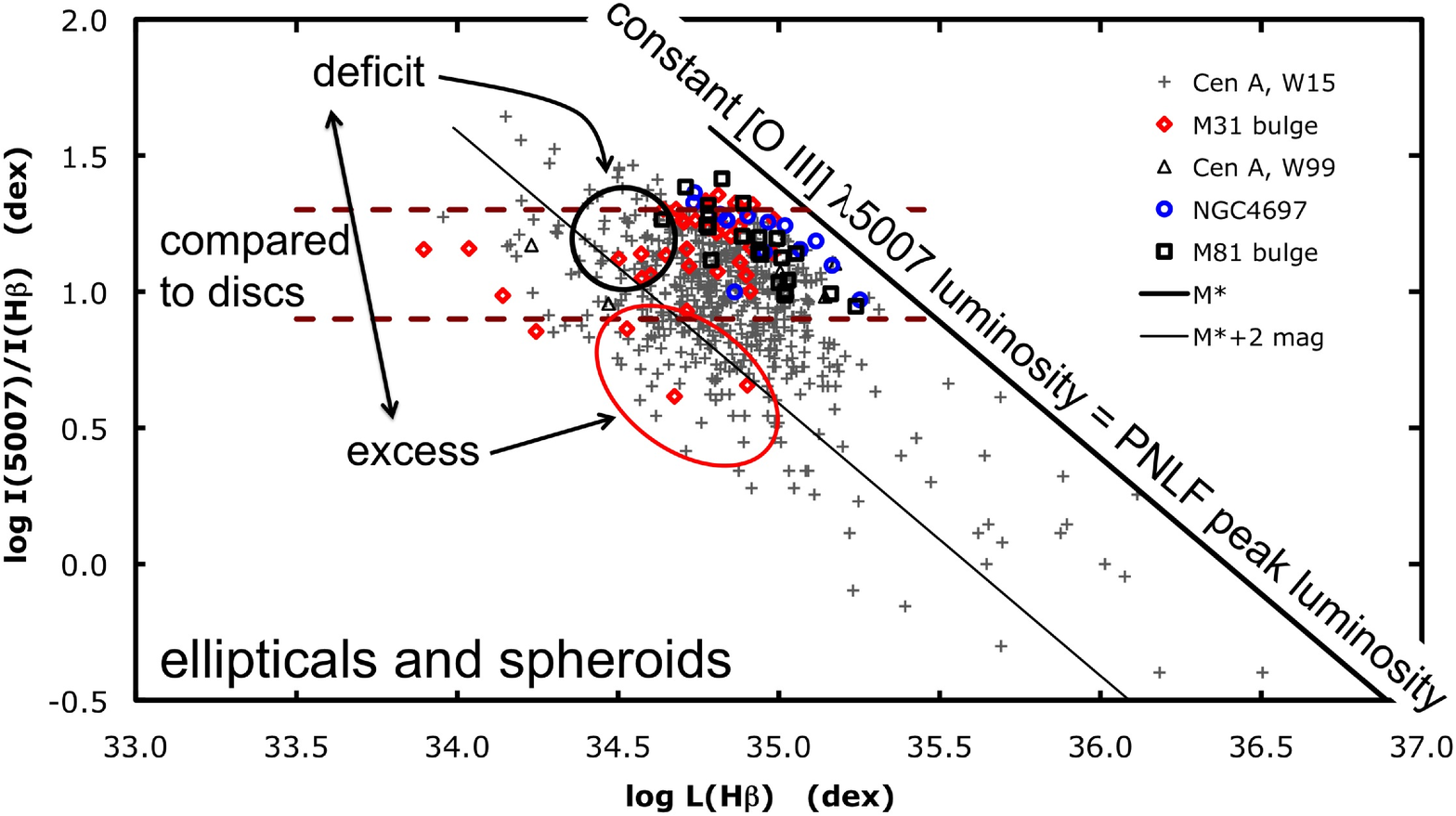} 
 \caption{These figures present the distribution of the ratio of [O {\sc iii}] $\lambda$5007/H$\beta$ as a function of the H$\beta$ luminosity for bright PNe in nearby galaxies, separated into spheroids and discs.  In both panels the bold diagonal line indicates the peak luminosity of the PNLF while the lighter diagonal line indicates an [O {\sc iii}] $\lambda$5007 luminosity 2 mag fainter.  In both panels, the horizontal lines indicate the band occupied by PNe in the discs of spirals.  The ellipses indicate where the fainter PNe dominate in both samples.  The circle indicates a lack of bright PNe in ellipticals and spheroids.  In general, the disc systems (left panel) are nearer than the ellipticals and spheroids, explaining the much greater range of H$\beta$ luminosities probed in disc systems.  In neither case are the samples statistically complete.  Even so, there are clear clues that the distributions of the brightest PNe differ between discs and spheroids.  Data sources:  Balick et al. (2013, ApJ, 774, 3); Bresolin et al. (2010, MNRAS, 404, 1679); Corradi et al. (2015, ApJ, 807, 181); Fang et al. (2013, ApJ, 774, 138); Jacoby \& Ciardullo (1999, ApJ, 515, 169); Jacoby \& Ford (1986, ApJ, 304, 490); Kniazev et al. (2014, AJ, 147, 16); Kwitter et al. (2012, ApJ, 753, 12); Magrini et al. (2009, ApJ, 696, 729); McCall (2014, MNRAS, 440, 405); M\'endez et al. (2005, ApJ, 627, 767); Richer et al. (1999, A\&ASS, 135, 203); Richer et al. (2008, ApJ, 689, 203); Richer et al. (2010, ApJ, 716, 857); Richer \& McCall (unpublished); Roth et al. (2004, ApJ, 603, 531); Sanders et al. (2012, ApJ, 758, 133); Stanghellini et al. (2010, A\&A, 521, A3); Stasi\'nska et al. (2013, A\&A, 552, A12); Walsh et al. (1999, A\&A, 346, 753); Walsh et al. (2015, A\&A, 574, A109).  
}
   \label{fig1}
\end{center}
\end{figure}

We can use observations of the H$\beta$ luminosities and 5007/H$\beta$ ratios for bright PNe in different galaxies to determine whether the details of the evolution are the same in all galaxies.  If bright PNe in all galaxies arise from stars with the same masses, metallicities, and ages, the distribution of H$\beta$ luminosities and 5007/H$\beta$ ratios should be the same.  In Fig. \ref{fig1}, we see that they appear to differ, at least between the discs of spiral galaxies and ellipticals and the bulges of spirals, even for the PNe with the highest luminosities.  First, the majority of objects more than 2 mag below the PNLF peak in disc systems (left panel) have larger 5007/H$\beta$ ratios than those in ellipticals and spheroids (compare the ellipses in both panels).  Though their [O {\sc iii}] $\lambda$5007 luminosities are similar, the H$\beta$ luminosities differ (those of the PNe in spheroids are brighter), so there is no reason they should be missing from observations of the discs of spirals, if they exist.  Second, compared to disc systems, the ellipticals and spheroids have a deficit of PNe with luminosities 1-2 mag below the PNLF peak luminosity, but with high 5007/H$\beta$ ratios (circle, right panel).  In summary, there must be a variety of ways to produce the intrinsically brightest PNe.  

These differences may be exploited to determine the origin of a PN population in some cases.  In the left panel of Fig. \ref{fig1}, the median values are shown for the four evolutionary phases of bright PNe from the Milky Way bulge, ordered as indicated, from Richer et al. (2008, 2010).  Clearly, the locus defined by these median values are compatible with the the distribution of PNe in discs
, perhaps evidence that the Milky Way's bulge evolved secularly from its disc.

\end{document}